\documentclass[12pt,oneside,final,a4paper]{article}
\usepackage{amsmath,amssymb,makeidx,amsfonts,latexsym,enumerate}

\title{`Quantum probabilities' as context depending probabilities.}
\author{Andrei Khrennikov\\
School of Mathematics and Systems Engineering\\
University of V\"axj\"o, S-35195, Sweden}
\begin{document}
\maketitle

\begin{abstract}
We study transformations of conventional (`classical') probabilities
induced by context transitions. It is demonstrated that the transition
from one complex of conditions to another induces a perturbation of
the classical rule for the addition of probabilistic alternatives. We classify 
such perturbations. It is shown that there are two classes of perturbations:
(a) trigonometric interference; (b) hyperbolic interference. In particular, the well
known `quantum interference of probabilistic alternatives' can be obtained in classical
(but contextual) probabilistic framework. Therefore we need not apply to wave arguments 
(or consider superposition principle) to get QUANTUM STATISTICS. In particular, interference
could be a feature of experiments with purely corpuscular objects.
\end{abstract}

\section{Introduction}
It is well known that the classical rule for the addition of probabilistic alternatives:
\begin{equation}
\label{F1}
P=P_1+P_2
\end{equation}
does not work in experiments with elementary particles. Instead of this rule, we have to use quantum rule:

\begin{equation}
\label{F2}
P=P_1+P_2+2\sqrt{P_1P_2}\cos\theta.
\end{equation}

So, the classical rule for the addition of probabilistic alternatives is perturbed by so called 
interference term. The difference between `classical' and `quantum' rules was (and is) the source of
permanent discussions and various misunderstandings (and mystifications). We just note that the appearance
of the interference term was the source of the wave-viewpoint to the theory of elementary particles;
at least the notion of superposition of quantum states was proposed as an attempt to explain the appearance
of a new probabilistic calculus in the two slit experiment, see, for example, Dirac's book [1] on historical 
analysis of the origin of quantum formalism. We also mention that Feynman interpreted (\ref{F2}) as the evidence
of the violation of the additivity postulate for `quantum probabilities', [2].

In particular, this induced the viewpoint that there are some special 
`quantum' probabilities that differs essentially from ordinary `classical' probabilities. We also remark that 
the orthodox Copenhagen interpretation of quantum formalism is just an attempt to explain (\ref{F2}) without 
to apply to mysterious `quantum probabilities'. To escape the use of a new probabilistic calculus, we could 
suppose that, e.g. electron participating in the two slit experiment is in the superposition of passing through
both slits.

\section{The role of a complex of conditions}
There is another possibility to explain the violation of ({\ref{F1}}) in experiments with elementary particles.
This is so called contextual explanation, see, for example, [3], [4], [5].
In fact, probabilities in, e.g., two slit experiment, 
are determined by three different contexts: ${\cal S}=${both slits are open}, ${\cal S}_j=${only jth slit is open}, $j=1,2.$ Here $P=P(A|{\cal S})$ and $P_j=P(A|{\cal S}_j), j=1,2.$ I think everybody would agree that the conventional probability theory does not say that $P(A|{\cal S})=P(A|{\cal S}_1)+P(A|{\cal S}_2).$
However, the contextual explanation is not so popular in quantum community. 
There are three main reasons for such unpopularity: a) psychology of scientists; 
b) the contextual approach contradicts 
to the conventional theory of probability based on Kolmogorov's axiomatics, [6];
c) the absence of the derivation of the interference perturbation of the `classical' rule, (\ref{F1}),
in the contextual framework (i.e., without to apply to Hilbert space tricks).

We could not do anything with (a). However, I recently discovered that, in fact, there is no contradiction 
between Kolmogorov's approach [6] and the contextual approach.
\footnote{I would like to thank Prof. A. Shiryaev (the former student of A. N. Kolmogorov) who explained this to me.}

It is the mysterious fact that a few generations of readers (including the author of this paper) 
did not pay attention to section 2 of Kolmogorov's book [6]. There Kolmogorov said directly 
that all probabilities of events are related to concrete complexes of conditions. Moreover, in his 
paper [7] he even used the symbol $P(A|{\cal S}),$ where ${\cal S}$ is a complex of conditions. 
This context dependence of probabilities
was so evident for Kolmogorov that he omitted the symbol ${\cal S}$ and used the symbol $P(A)$ 
to denote the probability. Unfortunately, it was (and is) not evident for everybody. Manipulation with 
`context less' probabilities became the standard tool of physical investigations (in particular, in 
experiments with elementary particles). Therefore you now need to cretize (in fact, without any reason) 
Kolmogorov's theory to attract the attention to the contextual approach, see, e.g. [8]. 
Regarding to (c): we shall derive (\ref{F2}) in purely classical (but contextual) framework in 
the next section.

\section{Interference term as the measure of statistical deviations due to the context transition.}
The following simple considerations induce `quantum probabilistic transformation' (\ref{F2}) in the 
classical probabilistic framework. Let ${\cal S}$ and ${\cal S}^\prime$ be two different complexes of conditions.
Suppose that each of them can be represented as a disjoint union of two subcomplexes of conditions: 

${\cal S}={\cal S}_1\cup {\cal S}_2, {\cal S}^\prime={\cal S}_1^\prime\cup {\cal S}_2^\prime.$

We consider the transformation of probabilities induced by transition from one complex of conditions to 
another: 

${\cal S}\rightarrow {\cal S}^\prime.$ 

We start with the standard addition of probabilistic alternatives for the fixed complex of conditions
${\cal S}.$ Let $B$ be some event. Then by using the classical formula we get: 
\begin{equation}
\label{K}
P(B|{\cal S})=P(B|{\cal S}_1)+P(B|{\cal S}_2).
\end{equation}
We remark that in many experiments there is (at least formal) correspondence between 
contexts ${\cal S}_j$ and ${\cal S}_j^\prime$ ($j=1,2):$ ${\cal S}_j \to {\cal S}_j^\prime.$
In such a situation it is `very attractive' to identify these contexts and use, instead of (\ref{K}),
the rule:
\begin{equation}
\label{K1}
P(B|{\cal S})=P(B|{\cal S}_1^\prime)+P(B|{\cal S}_2^\prime).
\end{equation}
This is the main root of {\it quantum mystification.}

The trivial remark is that we can rewrite conventional rule (\ref{K}) as:
\begin{equation}
\label{F3}
P(B|{\cal S})=P(B|{\cal S}_1^\prime)+P(B|{\cal S}_2^\prime)+\delta({\cal S}, {\cal S}^\prime),
\end{equation}

where 

$\delta({\cal S}, {\cal S}^\prime)=[P(B|{\cal S}_1)-P(B|{\cal S}_1^\prime)]+[P(B|{\cal S}_2)-P(B|{\cal S}_2^\prime)].$

Transformation (\ref{F3}) is the most general form of probabilistic transformations due to context transitions. 
There is {\it{the correspondence principle}} between context unstable and (`classical') context
stable transformations: If ${\cal S}^\prime\rightarrow {\cal S},$ i.e., $\delta({\cal S}, {\cal S}^\prime) \rightarrow 0$, 
then contextual probabilistic transformation (\ref{F3}) coincides (in the limit) with the classical rule for the addition of probabilistic alternatives.

The perturbation term $\delta({\cal S}, {\cal S}^\prime)$ depends on absolute magnitudes of probabilities. 
It would be natural to introduce normalized coefficient of the context transition 

$\lambda({\cal S}, {\cal S}^\prime)=\frac{\delta({\cal S}, {\cal S}^\prime)}{2\sqrt{P(B|{\cal S}_1^\prime)P(B|{\cal S}_2^\prime)}},$

that gives the relative measure of statistical deviations due to the transition from one complex of conditions, ${\cal S},$ to another, ${\cal S}^\prime.$ Transformation (\ref{F3}) can be written in the form:

\begin{equation}
\label{F4}
P(B|{\cal S})=P(B|{\cal S}_1^\prime)+P(B|{\cal S}_2^\prime)+
2\sqrt{P(B|{\cal S}_1^\prime)P(B|{\cal S}_2^\prime)}\lambda_B({\cal S},{\cal S}^\prime)\;.
\end{equation}

Another trivial remark is that, in fact, there are two possibilities:

1). $|\lambda({\cal S}, {\cal S}^\prime)|\leq 1;$

2).$|\lambda({\cal S}, {\cal S}^\prime)|\geq 1.$

In the first case we can always represent the context transition coefficient in the form:
\[\lambda({\cal S}, {\cal S}^\prime)=\cos \theta ({\cal S}, {\cal S}^\prime);\] in the second case:

\[\lambda({\cal S}, {\cal S}^\prime)=\pm\cosh \theta ({\cal S}, {\cal S}^\prime).\]

We have two types of probabilistic transformations induced by the transition from one complex of conditions to another: 

\begin{equation}
\label{F5}
P(B|{\cal S})=P(B|{\cal S}_1^\prime)+P(B|{\cal S}_2^\prime)+2\sqrt{P(B|{\cal S}_1^\prime)P(B|{\cal S}_2^\prime)}\cos\theta,
\end{equation}
\begin{equation}
\label{F6}
P(B|{\cal S})=P(B|{\cal S}_1^\prime)+P(B|{\cal S}_2^\prime)+\pm\sqrt{P(B|{\cal S}_1^\prime)P(B|{\cal S}_2^\prime)}\cosh\theta.
\end{equation}

So, we get `quantum probabilistic rule', (\ref{F2}) in the classical probabilistic framework 
(in particular, without any reference to superposition of states) by taking into account context
dependence of probabilities. We underline that (\ref{F2})(=(\ref{F5})) describes context transitions of 
relatively small magnitudes: $|\lambda|\leq 1.$ Complex waves can be obtained by using the following 
formula: 

$a+b+2\sqrt{ab}\cos\theta=|\sqrt{a}+\sqrt{b}e^{i\theta}|^2, a,b >0.$ 

Thus the context transition 
${\cal S}\rightarrow {\cal S}^\prime$ can be described by the wave:

$\varphi=\sqrt{P(B|{\cal S}^\prime_1)} + \sqrt{P(B|{\cal S}_2^\prime)}e^{i\theta_B({\cal S}, {\cal S}^\prime)}.$

The next step forward to the Hilbert space formalism is to represent $\varphi$ as a superposition of 
two states with complex coefficients. Here we can follow to P. Dirac [1]. The main distinguishing feature 
of our considerations is the absence of any dependence of the theory on special properties of elementary 
particles. There can be easily presented experiments with macro objects that produce 
(due to changes of complexes of conditions) transformation (\ref{F5}). 

Relatively large statistical deviations are described by transformation (\ref{F6}). Such transformations do not 
appear in the conventional formalism of quantum mechanics. In principle, they could be described by so called 
{\it{hyperbolic quantum mechanics}}, [9]. However, it may be that transformations (\ref{F6}) could be described 
by some generalization of the conventional quantum formalism, e.g., theory based on 
{\it positively defined operator valued measures.}
Some variants of the contextual approach to quantum probabilities were presented in author's works [10]-[12]. 
I hope that in the present paper it was, finally, found the clear and simple presentation of those ideas:

For each fixed context (experimental arrangement), we have CLASSICAL STATISTICS.
Transition from one complex of conditions to another induces interference perturbation
of the conventional rule for the addition of probabilistic alternatives. However, 
interference addition of probabilities can be obtained in purely classical probabilistic
framework.

I would like to thank S. Albeverio, L. Accardy, L. Ballentine, W. De Myunck, T. Hida
and A. Shiryaev for fruitful discussions.

{\bf References}

[1] P. A. M.  Dirac, {\it The Principles of Quantum Mechanics.}
(Claredon Press, Oxford, 1995).

[2] R. Feynman and A. Hibbs, {\it Quantum Mechanics and Path Integrals}
(McGraw-Hill, New-York, 1965).

[3] L. E. Ballentine, {\it Am. J. Phys.}, {\bf 54,} 883 (1986).

[4] L. Accardi, {\it Urne e Camaleoni: Dialogo sulla realta,
le leggi del caso e la teoria quantistica.} Il Saggiatore, Rome (1997).

[5] W. De Muynck, W. De Baere, H. Marten,
{\it Found. of Physics,} {\bf 24}, 1589--1663 (1994).

[6] A. N. Kolmogoroff, {\it Grundbegriffe der Wahrscheinlichkeitsrechnung.}
Springer Verlag, Berlin (1933); reprinted:
{\it Foundations of the Probability Theory}. 
Chelsea Publ. Comp., New York (1956).

[7] A. N. Kolmogorov, {\it Theory of Probability.} In series {\it Mathematics, its context, methods and role},
{\bf 2}, 252-290, Academy of Sc. of USSR, Steklov Math. Institute (in Russian).

[8] A.Yu. Khrennikov, {\it Interpretations of 
probability} (VSP Int. Publ., Utrecht, 1999).
 
[9] A. Yu. Khrennikov, {\it Hyperbolic Quantum Mechanics.} Preprint: quant-ph/0101002, 31 Dec 2000.

[10] A. Yu. Khrennikov, {\it Ensemble fluctuations and the origin of quantum probabilistic
rule.} Rep. MSI, V\"axj\"o Univ., {\bf 90}, October (2000).

[11] A. Yu. Khrennikov, {\it Classification of transformations of  probabilities for preparation procedures:
trigonometric and hyperbolic behaviours.} Preprint quant-ph/0012141, 24 Dec 2000.

[12] A. Yu. Khrennikov, {\it Linear representations of probabilistic transformations induced
by context transitions.} Preprint quant-ph/0105059, 13 May 2001.

\end{document}